\documentclass[a4paper,11pt]{article}
\pdfoutput=1 

\usepackage{jcappub}
\usepackage[utf8]{inputenc}  

\usepackage[T1]{fontenc} 
\usepackage{float}
\usepackage{xcolor}

\usepackage{caption}
\usepackage{subcaption}

\title{\boldmath Non-local gravity in bouncing cosmology scenarios.}

\author[a]{D. Jackson,\note{Corresponding author.}}
\author[b,1]{R. Bufalo,}

\affiliation[a]{Instituto de Física Teórica, IFT-UNESP,\\R. Dr. Bento Teobaldo Ferraz, 271, Várzea da Barra Funda, São Paulo - SP, 01140-070, Brazil}
\affiliation[b]{Departamento de F\'isica, Universidade Federal de Lavras\\Caixa Postal 3037, 37200-900 Lavras, MG, Brazil}

\emailAdd{dimas.jackson@unesp.br}
\emailAdd{rodrigo.bufalo@ufla.br}

\abstract{
In this work, we analyzed the improved Deser-Woodard non-local gravity over the background of five different bouncing cosmologies, whose premise is avoid the initial singular state of the universe.
We developed the numerical solution for the non-local distortion function, which encompass the modifications to the Einstein-Hilbert action, using the reconstruction procedure and we have found that they have a viable cosmological solution.
Afterwards, we discussed the physical aspects and outcomes of the evolution of the distortion function throughout the bouncing point for these models, specifically: the symmetric bounce, oscillatory bounce, the matter bounce, finite time singularity model, and the pre-inflationary asymmetrical bounce.
}

\begin{document}
\maketitle
\flushbottom

\section{Introduction}
\label{sec:intro}

Despite all the success in comparison of observational data and
consistency of the theory in mathematical terms \cite{will2014confrontation}, the Einstein's theory of  General Relativity (GR) present some limitations.
A well-known example is the equivalence principle, which is not directly applicable to the initial instant of the universe (in a model of ``big bang'' type).
Furthermore, there is a break in the necessary conditions for its application due to a singularity of space-time, i. e., this initial state has density, pressure and temperatures approximately
infinite \cite{hawking1973large, misner1973gravitation}.
In fact, there is no consensus on how to define metric singularities in GR, unlike electromagnetism for example, in which a singularity of the electromagnetic field can be avoided by defining 
the field in all the space except the singular point.
In the GR there is not a background spacetime to make this type of definition, since it is the spacetime itself
which becomes singular \cite{geroch1968singularity}.

In order to address the aforementioned space-time singularity, 
we would like to bring attention  to the proposal of bouncing cosmologies 	\cite{Bojowald:2001xe,novello2008bouncing,Ashtekar:2011ni,battefeld2015critical,Brandenberger:2017ni,Nojiri:2017ni}, 
where the cosmic contraction reduces the effective radius
of the cosmos to a minimum size which then produces an
expanding Universe, replacing effectively the big bang singularity by a non-singular bouncing geometry in which all curvature invariants become finite.
Moreover, bouncing cosmologies have also been considered as strong candidates to the standard inflationary paradigm \cite{mukhanov:2005rb,odintsov:2015ni}.

The currently standard cosmological model, namely $\Lambda$-Cold Dark Matter (CDM), is based in the GR.
In this model the universe starts in a singular point and after the early epoch, the vacuum energy dominates driving the accelerated expansion.
The presence of unknown energy sources in the $ \Lambda$-CDM  model indicates the possibility that General Relativity is not enough to explain the aforementioned phenomena \cite{ishak2019testing}, thus, several modified gravitational theories
were proposed as attempts to generalize  Einstein's gravitational theory.
Usually, the most used approach for the description of such phenomena involve addition of new
(global and/or local) degrees of freedom, which can be described in terms of new fields or even by enforcing a symmetry principle.

Nowadays, an interesting point of view used to elaborate a modified gravitational theory
are those theories with presence of \textit{non-local effects},\footnote{For a detailed account regarding conceptual non-local aspects in field theories, gravity and cosmology, for instance causality, domain of validity, boundary conditions, etc, see ref.~\cite{belgacem2018nonlocal,Capozziello:2022lic} and the references therein for a review. } which emerged from
 the works of D. Dalvit \cite {dalvit1994running} and C. Wetterich \cite{wetterich1998effective}.
 Actually, Wetterich proposed adding a term proportional to $ R \square^{- 1} R $
 \footnote{ 
 It is worth mentioning that this kind of non-local term are induced in loop corrections of the quantum effective action \cite{Barvinsky:1985rb,Buchbinder:1992rb,Mukhanov:2007rb,Shapiro:2008sf}.}
to the Einstein-Hilbert action and analyzed the cosmological implications
of a simple model based on this non-local action.
He concluded that his model was not compatible with primordial nucleosynthesis, whereas
generalizations consistent with experimental observations would be
conceivable.
Among the many interesting proposals of non-local modifications to the gravitational field action
\cite{belgacem2018nonlocal, maggiore2014nonlocal, barvinsky2012serendipitous}, we may refer to the S. Deser and R. Woodard  non-local model, where they have developed a theory with more general non-local terms of the form $ R f \left(\square^{-1} R \right) $ \cite{deser2007nonlocal}.
They have found that this model contains the time interval for the transition from radiation dominated universe to the matter dominated one, but several issues were still open.
A decisive test for this model was that the screening mechanism to avoid non-local effects in Solar System scale is not satisfied, violating thus some experimental constraints \cite{belgacem2018nonlocal}.
Afterwards, Deser and Woodard improved their model \cite{deser2019nonlocal} in order to satisfy the screening mechanism and also to reproduce the late time accelerated expansion without the necessity of a cosmological constant.

Naturally, due to the rich phenomenological scenario presented by the bouncing cosmology, it is very interesting to explore whether modified gravitational models provide viable cosmological solution in this context, see \cite{caruana2020cosmological} and the references therein for a review.
Actually, the first Deser-Woodard non-local cosmological model provided a viable description of a bouncing universe, which can be smoothly connected to the accelerated expansion scenario \cite{chen2019primordial}. 
Hence, our proposal consists in examining the viability of bouncing cosmologies within the improved non-local Deser-Woodard framework where we consider the five most studied bouncing cosmology scenarios, namely symmetric bounce \cite{Cai:2012va,Cai:2013vm,Bamba:2014mya}, oscillatory bouncing cosmology \cite{Steinhardt:2001st,novello2008bouncing,Cai:2012va}, matter bounce \cite{Singh:2006im,Wilson-Ewing:2012lmx}, the singularity cases \cite{Barrow:2015ora,Nojiri:2015fra,Odintsov:2015zza,Oikonomou:2015qha}
and the pre-inflationary asymmetrical bounce \cite{odintsov2022preinflationary}.
The singularity cases includes the Big Rip singularities, sudden singularities and the Big Freeze universe, all generated by the same scale factor but with different parameters.

In the present work we analyze the description of bouncing universe within the improved non-local Deser-Woodard model.
This is achieved by considering a flat Friedmann-Lemaître-Robertson-Walker (FLRW) geometry in which the bouncing frameworks emerge through the choice of particular forms of the scale factor.
In Sec.~\ref{sec:imp} we review the main features of the improved Deser-Woodard model, in special, focusing in obtaining the equations of motion for the non-local distortion function and the auxiliary fields.
In Sec.~\ref{sec:ani} we determine the solutions of five different classes of bouncing: symmetric, oscillatory, matter, singular, and pre-inflationary asymmetrical, and analyze how they are physically viable cosmological scenarios.
At last, in Sec.~\ref{sec:non} we discuss the issues related with the singularity cases and establish the main points rendering a nonphysical cosmological evolution.
We present our final remarks and perspectives in Sec.~\ref{sec:conc}.

\section{Improved Deser-Woodard gravity }
\label{sec:imp}

The improved non-local Deser-Woodard model \cite{deser2019nonlocal} minimally modifies the Einstein-Hilbert action, introducing two non-local scalar fields, such that
\begin{equation}
\mathcal{L}=\frac{1}{16\pi G}R\left[1+f(Y)\right],\label{eq:LDW}
\end{equation}
with the following definitions
\begin{subequations}
\begin{align}
Y & =\square^{-1}g^{\mu\nu}\partial_{\mu}X\partial_{\nu}X\,, \\
X & =\square^{-1}R\,.\label{eq:X}
\end{align}
\end{subequations}
Here $\square=g^{\mu\nu}D_{\mu}D_{\nu}$ is the d'Alembertian operator and $R$ is the curvature scalar.
Along this paper we will refer to the model described by the Lagrangian density \eqref{eq:LDW} as DW II.
This model was originally proposed to reproduce the late time accelerated expansion of the universe, without the need of the cosmological constant, only via a dynamical mechanism inspired by quantum effective action corrections.
Actually, the term $g^{\mu\nu}\partial_{\mu}X\partial_{\nu}X$ is negative in the Solar System scale and positive in the cosmological scale, which makes viable a screening effect.
As a consequence of this mechanism, the non-local corrections appear only on the cosmological background and the GR is recovered on strongly bound gravitational systems.
Thus, DW II obey the well established Solar System constraints \cite{ding2019structure}

It is important to our purposes to emphasize that the (dynamical and auxiliary) fields in \eqref{eq:LDW} are subject to retarded boundary conditions \cite{deser2019nonlocal}, which require that all the fields and its first time derivatives vanish in a initial value surface (in our case, in the early time of the contraction phase). These conditions are known as the Cauchy boundary conditions, which specifies the value of both function and its derivative on the boundary of the domain. In summary, unless these boundary conditions are satisfied by all the auxiliary scalars fields, new degrees of freedom would arise and possible negative kinetic terms, known as ghosts \cite{deser2019nonlocal}.
These retarded boundary conditions will be used throughout our analysis.

To localize the action, one can introduce two auxiliary scalar fields $U$ and $V$ as Lagrange multipliers,
\begin{align}
\mathcal{L} & =\frac{1}{16\pi G}\left[R\left(1+f(Y)\right)+U\left(R-\square X\right)+V\left(g^{\mu\nu}\partial_{\mu}X\partial_{\nu}X-\square Y\right)\right]\,, 
\end{align}
or, equivalently
\begin{equation}\label{eq:localDW}
\mathcal{L}=\frac{1}{16\pi G}\left[R\left(1+f(Y)+U\right)+g^{\mu\nu}E_{\mu\nu}\right],
\end{equation}
where we have introduced $E_{\mu\nu} = \partial_{\mu}U\partial_{\nu}X+\partial_{\mu}V\partial_{\nu}Y+V\partial_{\mu}X\partial_{\nu}X$, by means of notation.
Hence, considering $X,Y,U$ and $V$ as four independent scalar fields, the action $S=\int d^4x\mathcal{L}$ is regarded as local.
Varying the action in respect to each of these fields, result into the following set of equations
\begin{subequations}
\label{eq:constraints}
\begin{align}
R-\square X &=0\,,\label{eq:XDWII}
\\
g^{\mu\nu}\partial_{\mu}X\partial_{\nu}X-\square Y &=0\,,\label{eq:Y}
\\
2D^{\mu}\left(VD_{\mu}X\right) + \square U &=0\,,\label{eq:U}
\\
R\frac{\partial f(Y)}{\partial Y} -\square V &=0\,.\label{eq:V}
\end{align}
\end{subequations}
The gravitational field equations are obtained by varying the action  \eqref{eq:localDW} in respect to the metric $g^{\mu\nu}$:
\begin{equation}
\left(G_{\mu\nu}-D_{\mu}D_{\nu}+g_{\mu\nu}\square\right)\left(1+U+f(Y)\right)+E_{(\mu\nu)}-\frac{1}{2}g_{\mu\nu}g^{\rho\sigma}E_{\rho\sigma}=8\pi GT_{\mu\nu}\,,\label{eq:DW2field}
\end{equation}
where the energy momentum tensor $T_{\mu\nu}=\left(\rho+p\right)u_{\mu}u_{\nu}+pg_{\mu\nu}$ corresponds to the usual baryonic matter and does not include the dark energy source term \footnote{The indices in parenthesis denote the symmetric part $
E_{(\mu\nu)}=\frac{1}{2}\left(E_{\mu\nu}+E_{\nu\mu}\right)\,.$}.
 
This model reproduces the current accelerated expansion  of the universe without cosmological constant for 
 \begin{equation}
 	f(Y) \sim e^{1.1(Y+16.7)} \,.
 \end{equation}
 This is an exponential fit to the numerical solution obtained trough the reconstruction process \cite{deser2019nonlocal}. The function $f(Y)$ is named \textit{non-local distortion function} and the procedure of reconstruction consists in requiring that the Friedmann equations of General Relativity should be satisfied by the DW II. 
 
 To perform this procedure, the field equations \eqref{eq:DW2field} have to be expanded over the Friedmann-Lemaître-Robertson-Walker (FLRW) background:
 \begin{equation}
	ds^2=dt^2-a(t)dx_i dx^i\,.
 \label{eq:metric}
 \end{equation}
The d'Alembertian operator acting on a scalar function $W(t)$, which depends only on time, is written as
 \begin{align}
 \square W(t) &= d_{t}^{2}W(t)+3H d_{t}W(t)\,,\label{eq:BOXW}
 \end{align}
 where $H=\dfrac{\dot{a}}{a}$ is the Hubble parameter. In this background, the (00) and (ij) components of the modified field equations \eqref{eq:DW2field} become
 \begin{subequations}
 \begin{align}
 \left(3H^{2}+3Hd_t\right)\left(1+U+f(Y)\right)+\frac{1}{2}\left(\dot{X}\dot{U}+\dot{Y}\dot{V}+V\dot{X}^{2}\right) & =8\pi G\rho\,,\label{eq:00}
 \\
 -\left[2\dot{H}+3H^{2}+d_t^{2}+2Hd_t\right]\left(1+U+f(Y)\right)+\frac{1}{2}\left(\dot{X}\dot{U}+\dot{Y}\dot{V}+V\dot{X}^{2}\right) & =8\pi Gp\,.\label{eq:11}
 \end{align}
 \end{subequations}
 Furthermore, subtracting the equations \eqref{eq:00} and \eqref{eq:11} we find a differential equation for the function $F(t)\equiv1+U(t)+f(Y(t)$:
 \begin{equation}
 \left[2\dot{H}+6H^{2}+d_t^{2}+5Hd_t\right]F(t)=8\pi G\left(\rho-p\right)\,.\label{eq:FNL}
 \end{equation}
 Hence, the non-local distortion function can be numerically obtained through the relation $f=F-U-1$. We will perform this analysis in the next sections for some different bouncing universe solutions. All the development performed so far is independent of the particular form of the scalar factor $a(t)$, the  obtained field equations are satisfied by the generic FLRW metric in \eqref{eq:metric}.
In what follows, we will specify the form of the scale factor which reproduces different bouncing scenarios.

\section{Application to bouncing models}
\label{sec:ani}

In this section we will discuss the application of the model DW II to different bouncing models.
In special, we shall focus in the numerical analysis of the viability of the cosmological evolution near the bouncing point ($t=0$) of five different bounce classes: symmetric bounce, matter bounce, sudden singularity, finite time singularity model and the pre-inflationary asymmetric bounce. Furthermore, these bouncing models are illustrated in Figure \ref{fig:three graphs} by each scale factor.

\subsection{Symmetric bounce}
\label{sec:sym}

\begin{figure}[tbp]
	\centering
	\begin{subfigure}[b]{0.3\textwidth}
		\centering
		\includegraphics[width=\textwidth]{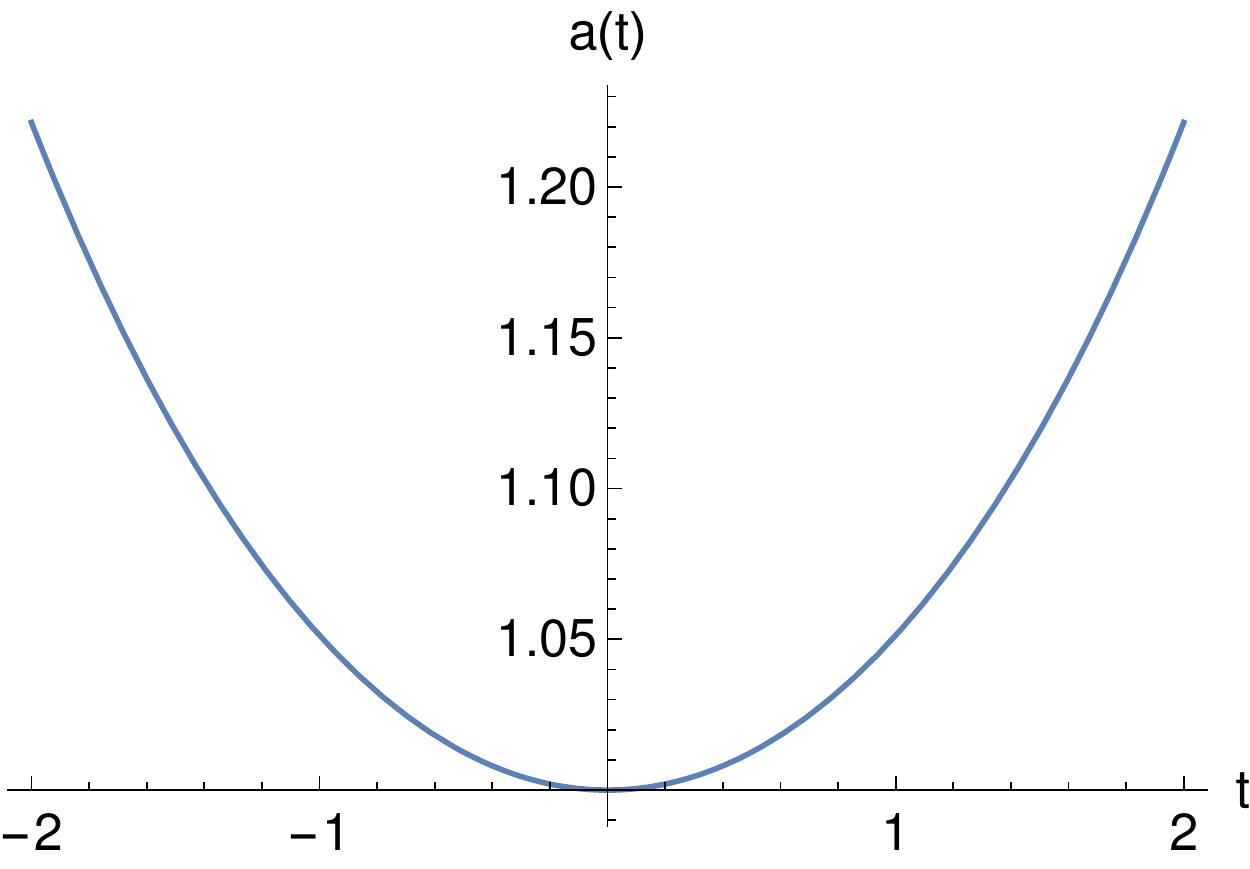}
		\caption{$a(t)=e^{t^{2}/20}\,.$}
		\label{fig:aI}
	\end{subfigure}
	\hfill
	\begin{subfigure}[b]{0.3\textwidth}
		\centering
		\includegraphics[width=\textwidth]{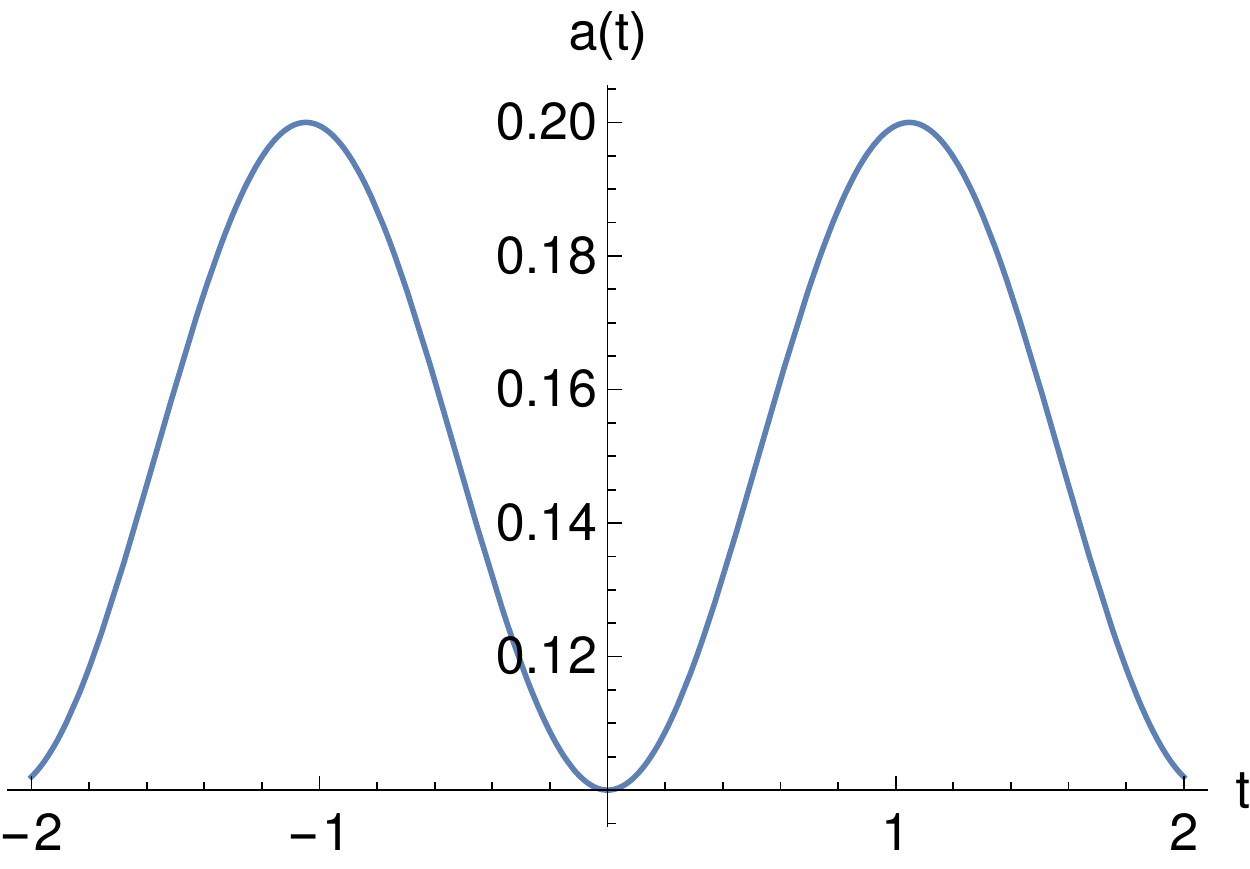}
		\caption{$a(t) = \frac{1}{10}[\sin^2{(\frac{3t}{2})}+1]$.}
		\label{fig:aII}
	\end{subfigure}
	\hfill
	\begin{subfigure}[b]{0.3\textwidth}
		\centering
		\includegraphics[width=\textwidth]{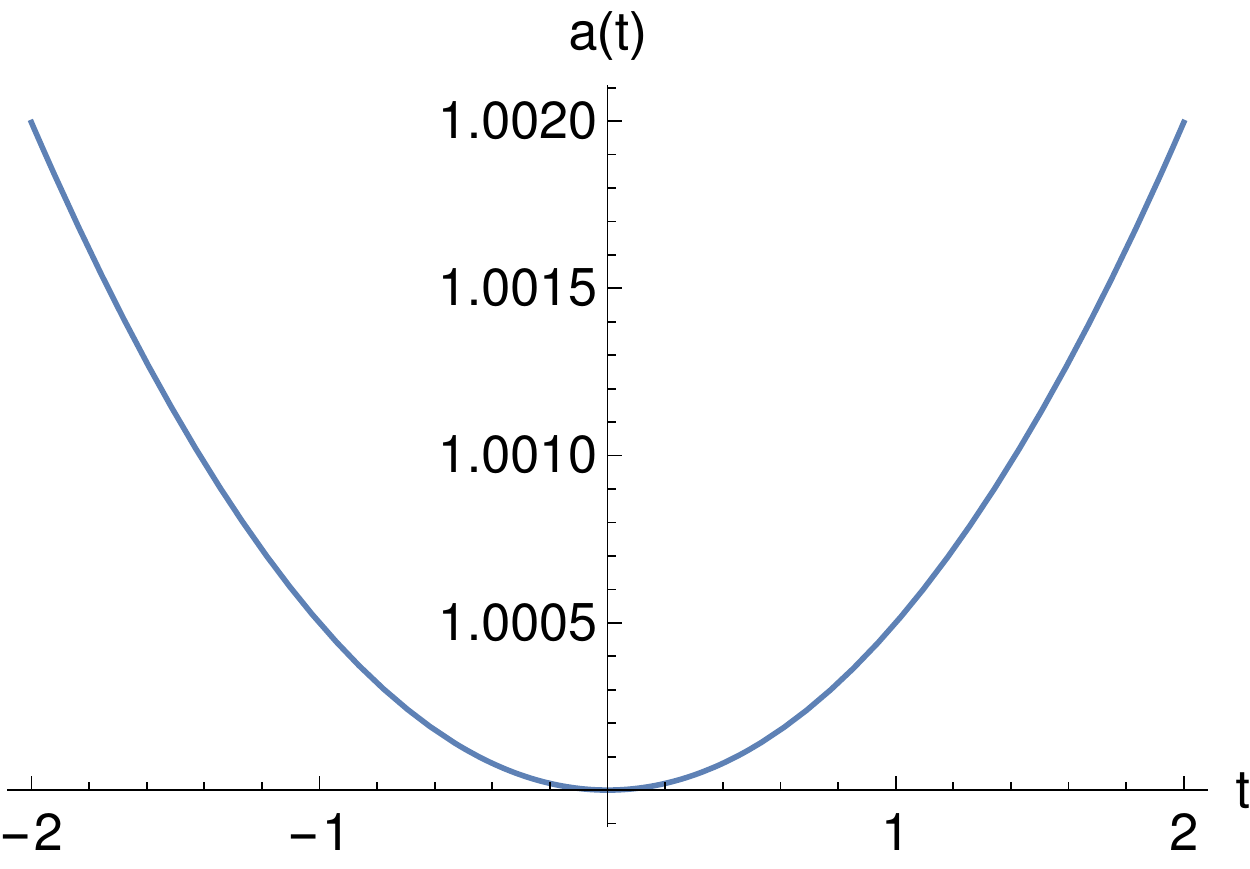}
		\caption{$a(t)=\left( \frac{3}{200}t^2+1 \right)^{1/3}$.}
		\label{fig:aIII}
	\end{subfigure}
		\hfill
	\begin{subfigure}[b]{0.3\textwidth}
		\centering
		\includegraphics[width=\textwidth]{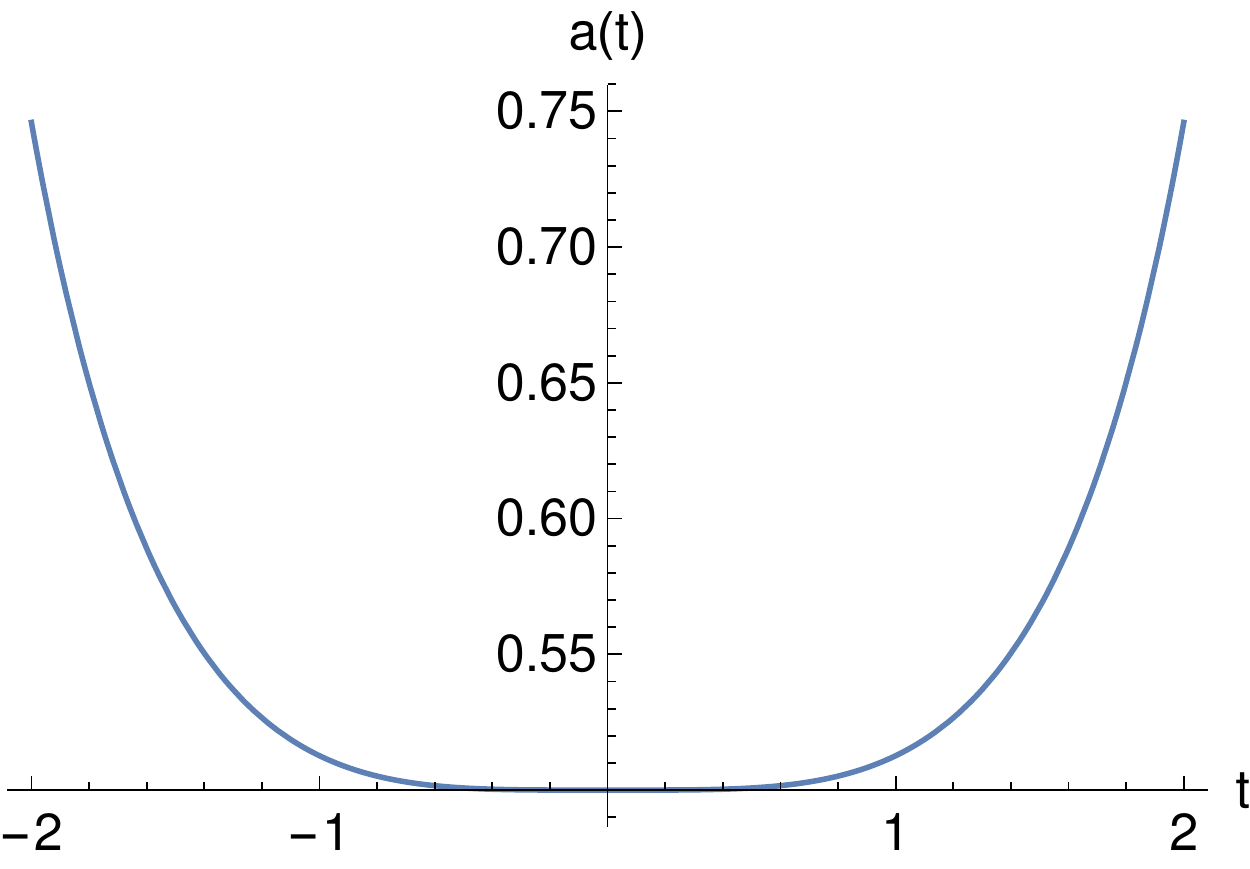}
		\caption{$a(t)=\frac{1}{2} e^{t^4/40}$.}
		\label{fig:aIV}
	\end{subfigure}
	\begin{subfigure}[b]{0.3\textwidth}
		\centering
		\includegraphics[width=\textwidth]{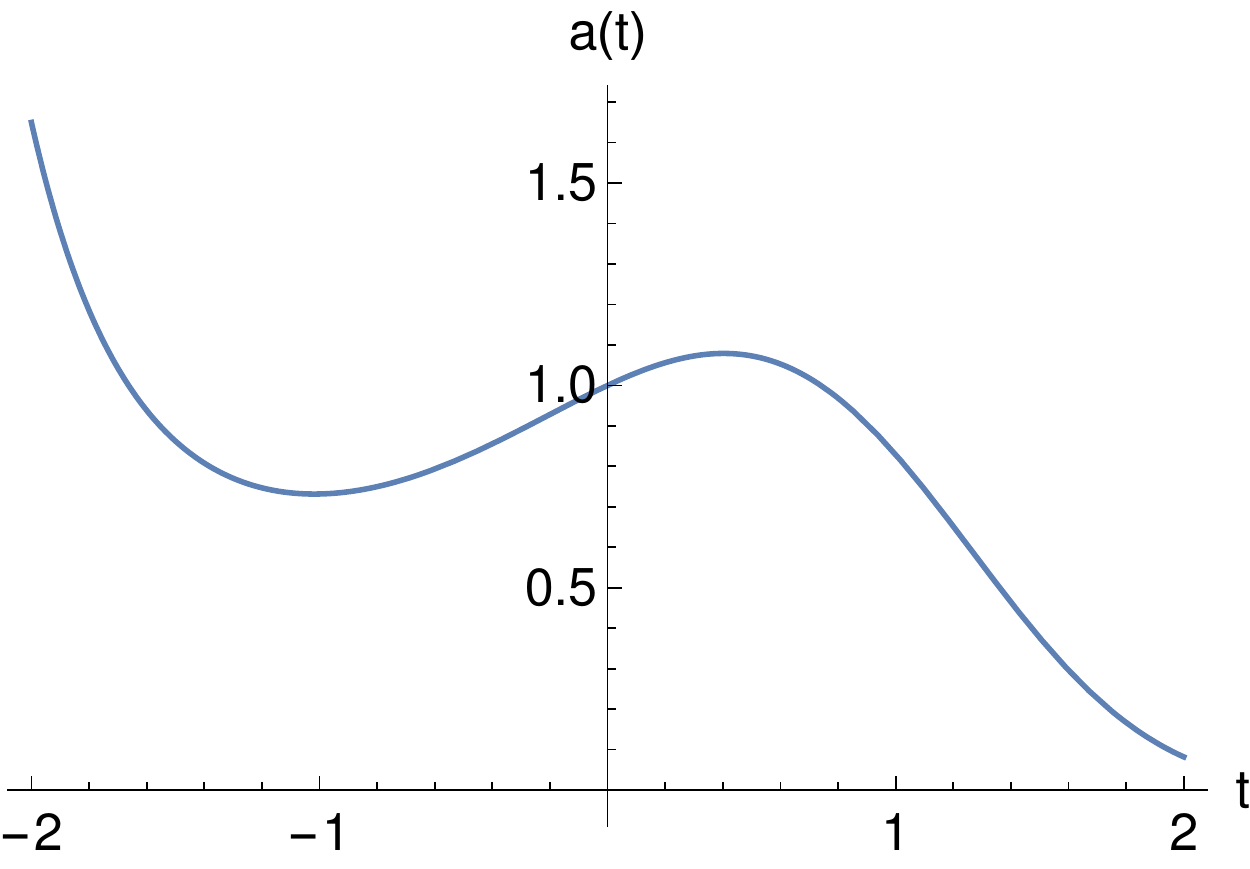}
		\caption{$a(t)=e^{-(\frac{11t}{17})^3-(\frac{t}{2})^2+\frac{t}{3}}$.}
		\label{fig:aV}
	\end{subfigure}
	\caption{The scale factor $a(t)$ for five different bouncing models: symmetric bounce, oscillatory bounce, matter bounce, finite time singularity model and pre-inflationary asymmetric bounce, respectively.}
	\label{fig:three graphs}
\end{figure}

We start our analysis of the DW II gravity in terms of the bouncing solutions by the case of the symmetric bounce, which generates a non-singular bounce and can be connected to the late time accelerated expansion. The symmetric bounce model also was studied in the context of the non-local DW I gravity \cite{chen2019primordial}, where it was considered as a viable cosmological scenario, and recently in the $f(T,B)$ gravity \cite{caruana2020cosmological}.
The symmetric bounce is characterized by the following scale factor \cite{Cai:2012va,Cai:2013vm,Bamba:2014mya,caruana2020cosmological}
\begin{equation}
a(t)=a_{m}e^{h_{1}t^{2}/2},
\label{eq:aI}
\end{equation}
where $a_{m}$ is the minimum value of scale factor and $h_{1}$ is
a constant, such that the Hubble parameter is $H(t)=h_{1}t$.
This scale factor is shown in panel (a) of the Figure \ref{fig:three graphs} and it is not derived from any equation of state, instead, it was postulated because of its desired behavior mentioned early.
We search for vacuum solutions of the equation \eqref{eq:FNL} i.e. $\rho=p=0$, so that
\begin{equation}
\left[2\dot{H}+6H^{2}+d_t^{2}+5Hd_t\right]F(t)=0\,.
\label{eq:dif}
\end{equation}
In the symmetrical bounce class, the differential equation (\ref{eq:dif}) has an exact solution, given by 
\begin{equation}
	F(t) = A_1 e^{-h_1t^2}+A_2H_{-1}\left( \sqrt{\dfrac{h_1}{2}}t \right)\,,
	\label{eq:FI}
\end{equation}
where $H_{n}$ is the Hermite polynomial (with negative index) and $A_{1},A_{2}$ are constants of integration. Using the relation
\begin{equation}
H_{-1}(x) = \dfrac{\sqrt{\pi}}{2}e^{x^2}\text{erfc}(x)\,,
\end{equation}
where $\text{erfc}\left(x\right) \equiv 1-\text{erf}(x)=1-\frac{2}{\sqrt{\pi}}\int_{0}^{x}e^{-u^{2}}du$
is the Gauss's complementary error function, the solution for $F(t)$ in eq.~\eqref{eq:FI} can be written in a suitable form as
\begin{equation}
F(t)=e^{-h_1t^{2}}\left(A_{1}+A_{2}\text{erfc}\left(\sqrt{\dfrac{h_{1}}{2}}t\right)\right)\,.
	\label{eq:F}
\end{equation}

Finally, in order to obtain the non-local distortion function $f=1-U-F$, it is necessary to find first the solutions for the fields $X,Y,U$ and $V$.
Actually, one can obtain a differential equation for the scalar field $ X $ by applying the background FLRW metric \eqref{eq:BOXW} to the constraint $ \square X = R $, eq.~\eqref{eq:XDWII}:
\begin{equation}
\left(d_t^{2}+3Hd_t\right)X+6\left(\dot{H}+2H^{2}\right)=0\,.
\end{equation}
Its exact solution is given in terms of well known special functions:
\begin{equation}
X(t)=B_{1}+B_{2}\text{erf}\left(\sqrt{\dfrac{3h_{1}}{2}}t\right)-h_{1}t^{2}\left(2+{}_2F_2\left(1,1;\dfrac{3}{2},2;-\dfrac{3}{2}h_{1}t^{2}\right)\right)\,,
	\label{eq:XIex}
\end{equation}
where ${}_2F_2\left(1,1;\dfrac{3}{2},2;-\dfrac{3}{2}h_{1}t^{2}\right)$ is the generalized hyper-geometric function and $ B_{1}, B_{2} $ are constants of integration.
Next, we can make use of the eq. \eqref{eq:Y} to write an differential equation for $Y$ in terms of $X$ as
\begin{equation}
\left(d_t^{2}+3Hd_t\right)Y-\dot{X}^{2}=0\,,
\end{equation}
but in this case, by making use of the solution eq.~\eqref{eq:XIex}, we can determine only numerical solutions for $Y$. Furthermore, from the definition of $U$ given in \eqref{eq:U}, over the background metric
FLRW, we have
\begin{equation}
\left(d_t+3H\right)\dot{U}=-2\left(d_t+3H\right)V\dot{X}\,,
\end{equation}
which can be equivalently expressed as the following relation
\begin{equation}
-2V=\dfrac{\dot{U}}{\dot{X}}\,.\label{eq:VUX}
\end{equation}
At last, over the background metric FLRW, the expression for $V$ in \eqref{eq:V} becomes
\begin{equation}
\left(d_t^{2}+3Hd_t\right)V=-6\left(\dot{H}+2H^{2}\right)\dfrac{df}{dY}\,. \label{eq_555}
\end{equation}
However, since the function $F=1+f+U$ was calculated in (\ref{eq:F}), then, we can eliminate $f$ by
\begin{equation}
\dfrac{df}{dY}=\dfrac{\dot{F}-\dot{U}}{\dot{Y}}\,. \label{eq_556}
\end{equation}
Thus, replacing \eqref{eq_556} back into \eqref{eq_555}, it results in the following 
\begin{equation}
\left(d_t^{2}+3Hd_t\right)V+6\left(\dot{H}+2H^{2}\right)\left(\dfrac{\dot{F}-\dot{U}}{\dot{Y}}\right)=0\,.
\end{equation}
It is important to emphasize that this step imposes the constraint $\dot{Y} \neq 0$ for all $t$ in the solution's interval, otherwise $V$, $U$ and hence $f$ will diverge everywhere.
Hence, substituting $\dot{U}$ from equation \eqref{eq:VUX}, one can finally obtain a differential equation for the auxiliary field $V$:
\begin{equation}
\left(d_t^{2}+3Hd_t\right)V+6\left(\dot{H}+2H^{2}\right)\left(\dfrac{\dot{F}+2V\dot{X}}{\dot{Y}}\right)=0\,.
\label{eq:EDOV}
\end{equation}

The numerical solution for the field $V$ provides the  necessary information to perform the numerical calculation of $U$ through equation \eqref{eq:VUX}.
Henceforth, the non-local distortion function $f=F-U-1$ is constructed from the numerical solution for $U$ and the exact equation \eqref{eq:F} for $F$.
The boundary conditions used to perform these calculations are the retarded boundary conditions, those previously mentioned in order to avoid the presence of (ghost) new degrees of freedom \cite{deser2019nonlocal}.
These conditions must also be satisfied by the function $F(t)$.

\begin{figure}[h]
	\centering 
\begin{subfigure}[b]{0.45\textwidth}
	\centering
	\includegraphics[width=\textwidth,clip]{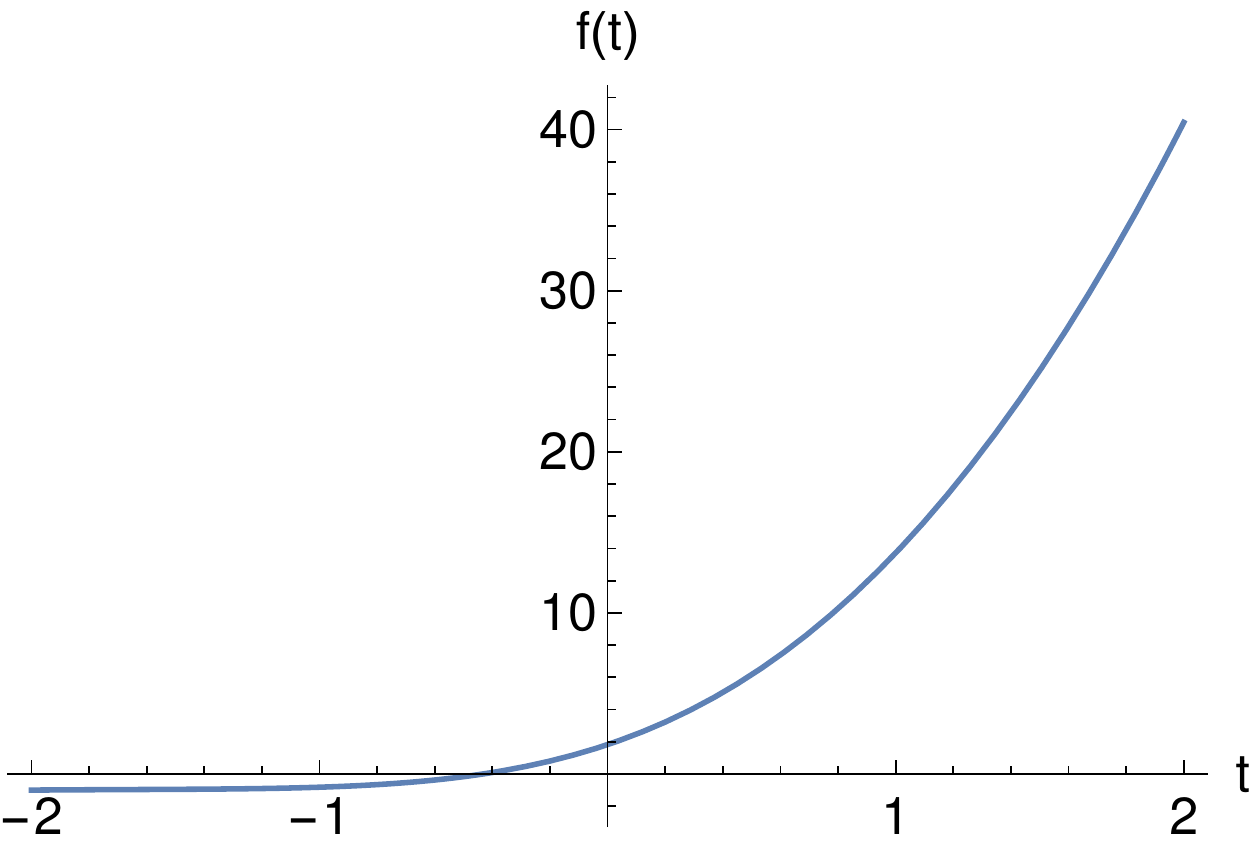}
	\caption{}
\end{subfigure}
	\hfill
\begin{subfigure}[b]{0.45\textwidth}
	\centering
	\includegraphics[width=\textwidth,clip]{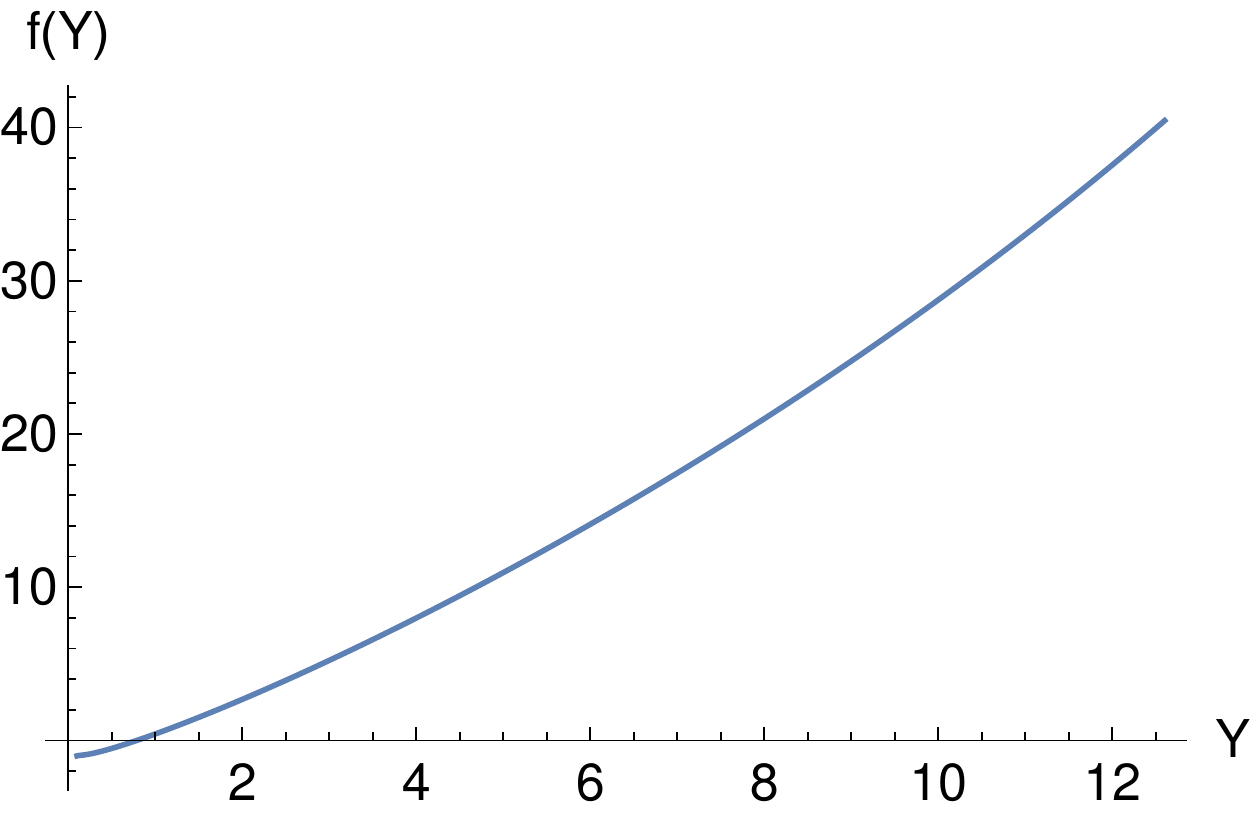}
	\caption{}
\end{subfigure}
	\caption{\label{fig:ftI} Non-local distortion function in the context of the symmetric bounce, panel (a) in terms of the cosmological time and (b) in terms of the non-local field $Y$.}
\end{figure}

The numerical solution for the distortion function in the symmetric bounce case is presented in panel (a) of the Figure \ref{fig:ftI}.
In this plot we can see the evolution of the distortion function throughout the collapse and rebirth of the universe.
In the symmetric bounce, the function $f(t)$ is an increasing function and assume a small value near the bouncing point. Observe that this function grows fast in the expanding phase.
 	
Finally, the solutions $f(t)$ and $Y(t)$ are used to obtain $f(Y)$. The numerical solution for $f(Y)$ is depicted in panel (b) of the Figure \ref{fig:ftI} and it corresponds to the non-local correction to Einstein-Hilbert action, as introduced by equation \eqref{eq:LDW}. As we can see, in this case, the non-local correction is also an increasing function and passes throughout the bouncing point ($Y = 2$) without changing its behavior. 

\subsection{Oscillatory bounce}

The oscillatory bounce, which we will discuss in this section, is generated by the following scale factor:
\begin{eqnarray}
a(t) = A\sin^2{(kt)}+c\,,
\end{eqnarray}
where $A$ and $k$ are positive constants that describes the bouncing frequency and the amplitude of the cyclic universe, respectively (see panel (b) of the Figure \ref{fig:three graphs}).
The parameter $c$ is also positive and is inserted to avoid the singularity, when the universe reaches its minimum size.
This model surges from the quasi-stead state cosmology \cite{sachs1996thequasi, novello2008bouncing}, which was proposed as an alternative to the standard cosmology.
The oscillatory pattern of the scale factor was introduced to reproduce the cyclic behavior of the interchange between the domination of the cosmological constant and a scalar field with negative energy that create particles.
In addition, similar models have been studied in the context of non-linear massive gravity and brane models \cite{cai2012bounce,mukherjo2002bouncing}.

\begin{figure}[H]
	\centering 
	\begin{subfigure}[b]{0.45\textwidth}
		\centering
		\includegraphics[width=\textwidth,clip]{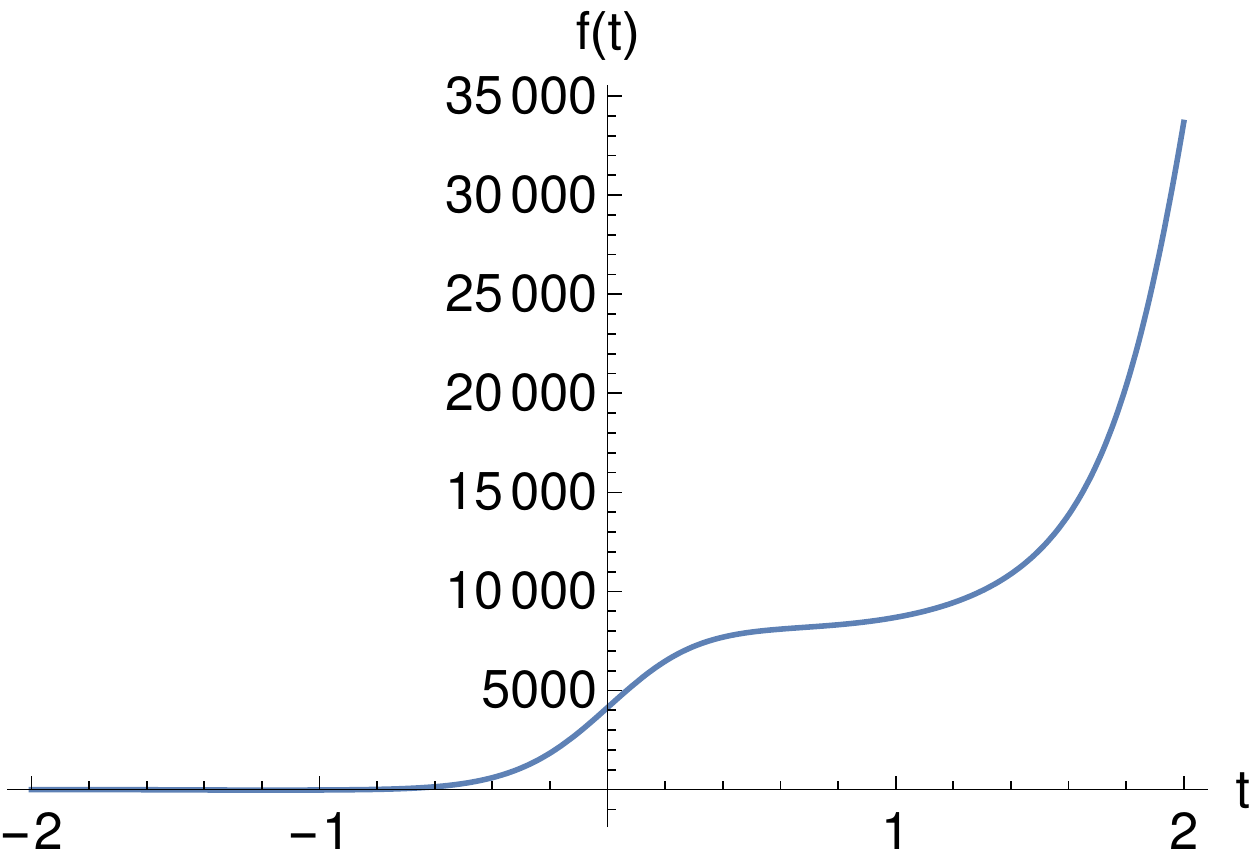}
		\caption{}
	\end{subfigure}
	\hfill
	\begin{subfigure}[b]{0.45\textwidth}
		\centering
		\includegraphics[width=\textwidth,clip]{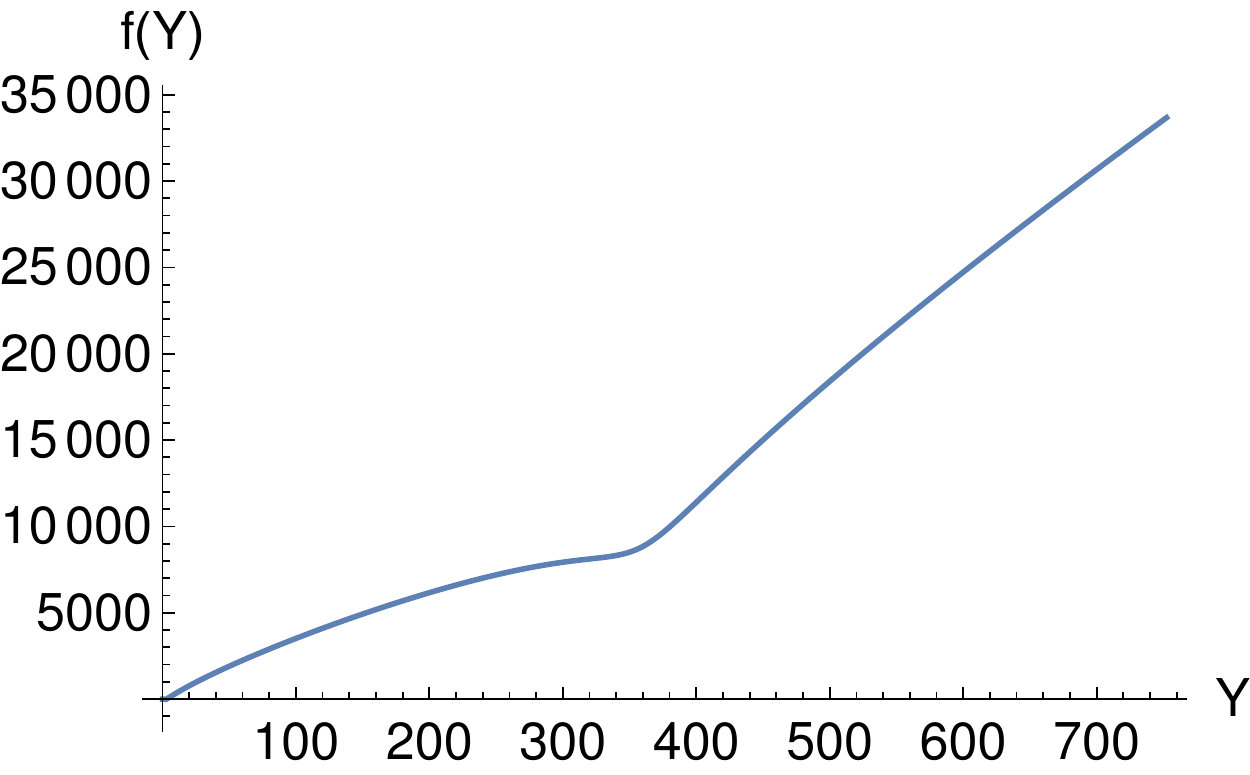}
		\caption{}
	\end{subfigure}
	\caption{\label{fig:ftII} Non-local distortion function in the context of the oscillatory bounce, panel (a) in terms of the cosmological time and (b) in terms of the non-local field $Y$.}
\end{figure}

We have reconstructed the non-local distortion function for this model using the same procedure described in the earlier section.
Although, in this case we have only found numerical solutions for the auxiliary fields $F,X,Y,U$ and $V$ and the same for the functions $f(t)$ and $f(Y)$.
Our results are presented in Figure \ref{fig:ftII}, see the panel (a) for the numerical solution for $f(t)$ and panel (b) for the distortion function.
In this scenario, $f(t)$ increases approximately linearly near the bouncing time ($t=0$), then becomes a plateau and finally grows to infinity when the next minimum size (Big Crunch) is reached.
The function $f(Y)$ is a continuous increasing function, but the growing rate starts to slow down near the principal bouncing, which occurs at $Y=200$.

\label{sec:osc}


\subsection{Matter bounce}

In this section we analyze the matter bounce, which emerged in the context of loop quantum cosmology (LQC) \cite{Singh:2006im}. This scenario is described by the following scale factor \cite{Singh:2006im,Wilson-Ewing:2012lmx,caruana2020cosmological}
\begin{eqnarray}
a(t) = A\left( \dfrac{3}{2}\rho _c t^2+1 \right)^{1/3}\,,
\end{eqnarray}
where $A$ is a constant and $\rho _c \ll 1$ is the critical density of the universe.
This scale factor satisfies the effective equations of LQC in the classical limit, for a dust-dominated universe.
These effective equations takes into account corrections due to quantum geometry in the usual Friedmann equations of the general relativity.

The calculation of the non-local distortion function $f(Y)$ follows the same development as presented in the previous section. Nonetheless, in this class of bounce, the exact solution for the function $F(t)$ is given in terms of Hyper-geometric functions:
\begin{eqnarray}
	F(t) = \frac{c_2 t  }{\sqrt[3]{2} \left(3 \rho _c  t^2+2\right)^{2/3}} \,_2F_1\left(\frac{1}{3},\frac{1}{2};\frac{3}{2};-\frac{1}{2} \left(3 t^2 \rho _c \right)\right) +\frac{c_1}{\left(3 \rho _c  t^2+2\right)^{2/3}}\,.
\end{eqnarray}
Furthermore, we also find an exact solution for the scalar field $X$, such that,
\begin{eqnarray}
X(t) = \frac{c_3 \tan ^{-1}\left(\sqrt{\frac{3}{2}} \sqrt{\rho _c } t\right)}{\sqrt{6} \sqrt{\rho _c }}+c_4-\frac{2}{3} \ln \left(3 \rho _c  t^2+2\right)-\frac{4}{3} \tan ^{-1}\left(\sqrt{\frac{3}{2}} \sqrt{\rho _c } t\right)^2\,.
\end{eqnarray}
The exact solution for field  $Y(t)$ is a huge expression, which does not highlight any important physical aspect, so we will not present it here.

\begin{figure}[h]
	\centering 
	\begin{subfigure}[b]{0.45\textwidth}
		\centering
		\includegraphics[width=\textwidth,clip]{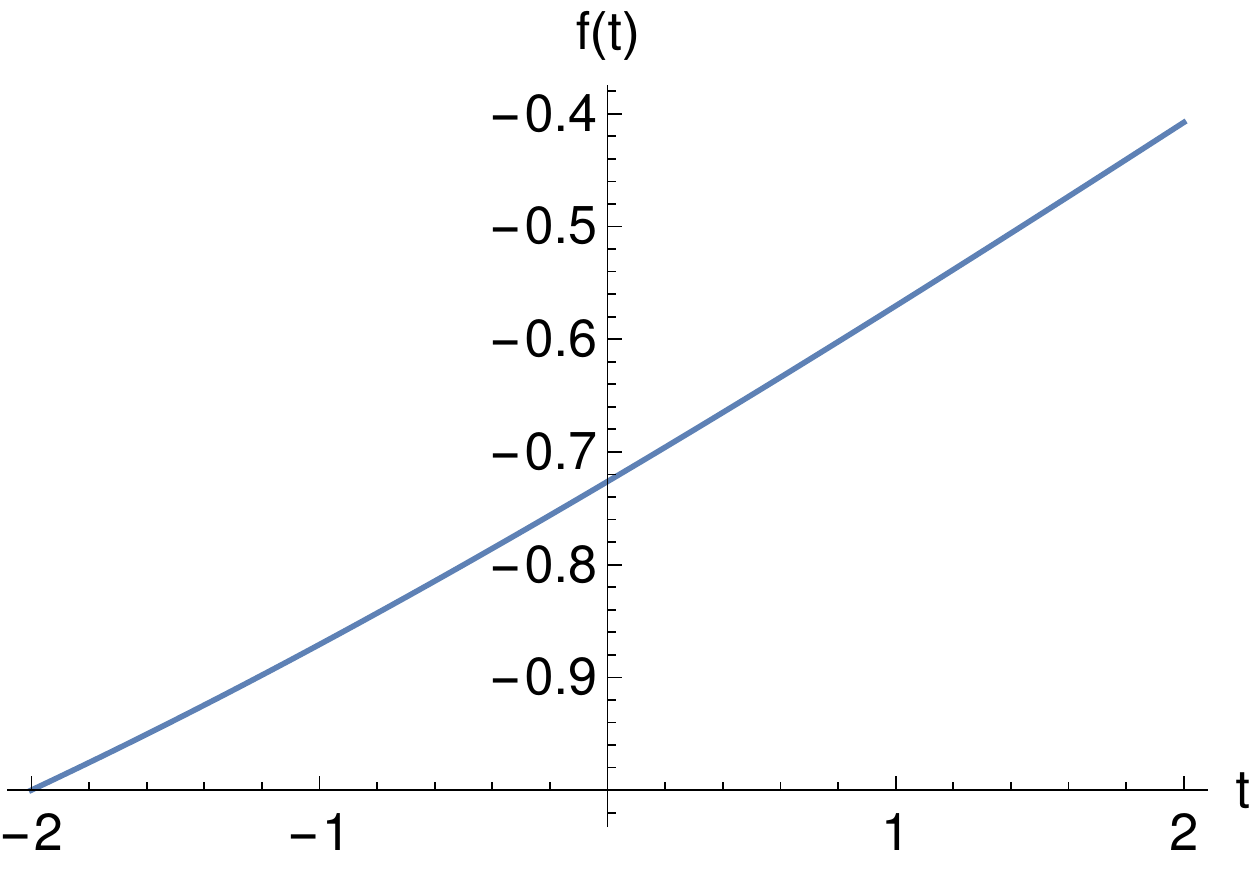}
		\caption{}
	\end{subfigure}
	\hfill
	\begin{subfigure}[b]{0.45\textwidth}
		\centering
		\includegraphics[width=\textwidth,clip]{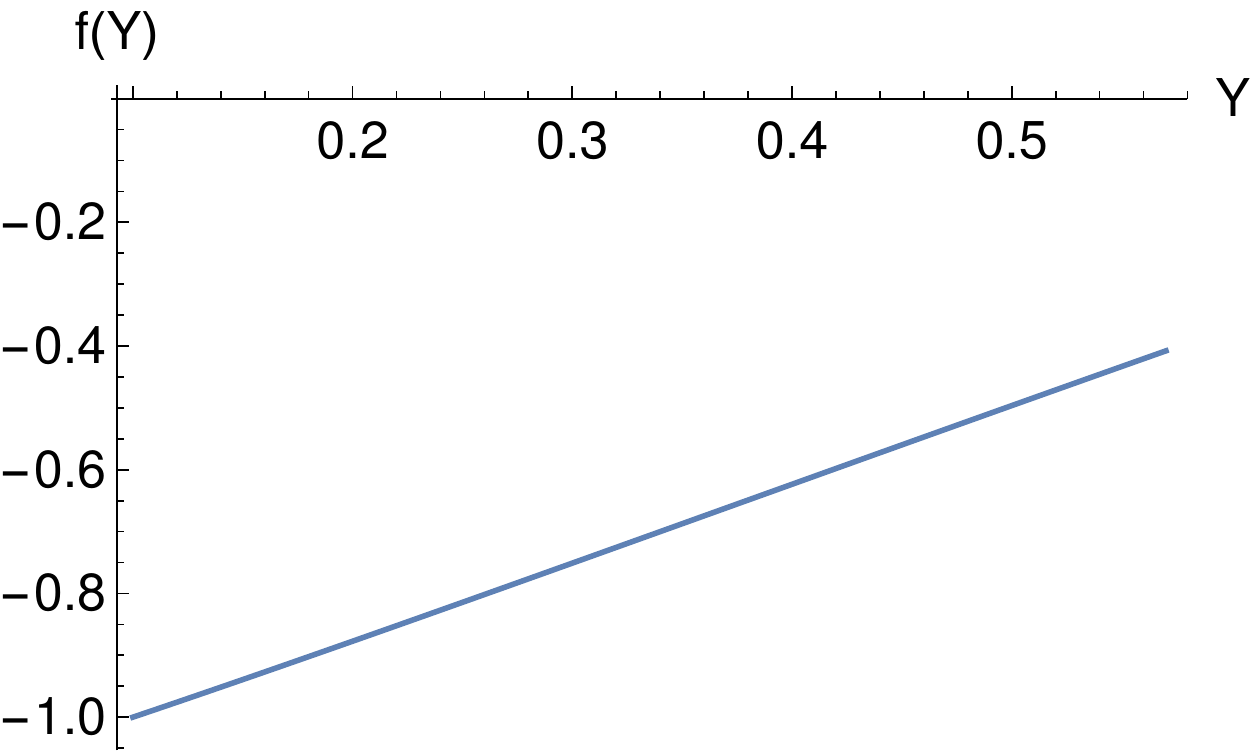}
		\caption{}
	\end{subfigure}
	\caption{\label{fig:ftIII} Numerical solution of the non-local distortion function in the matter bounce model, panel (a) in terms of the cosmological time and panel (b) in terms of the field $Y$.}
\end{figure}

On the other hand, the field equations for $U$ and $V$  had to be numerically solved, and were used to reconstruct the non-local function $f(t) = 1-U(t)-F(t)$.
The solution for $f(t)$ in the matter bounce scenario is a continuous function and is depicted in panel (a) of Figure \ref{fig:ftIII}. Note that the curve is smooth throughout the $t=0$ point.

Exploring the one-to-one relation between $Y$ and $t$, follow the numerical solution for $f(Y)$ depicted in panel (b) of Figure \ref{fig:ftIII}. We observe that, in this case, the distortion function is continuous from the past era to the present one, increases linearly and is strictly negative. The interval of solution used here $t \in [-2,2]$, can be extended as much as wished, obviously requiring more computational power. Therefore, the growth of the non-local distortion function at the endpoint, should be seen as occurring in the late time universe.
 

\subsection{Finite time singularity model}
\label{sec:big}
In this section we will analyze the behavior of the DW II gravity in the context of a more general exponential bouncing than that described in Section \ref{sec:sym}, which was originally proposed to discuss finite time singularities \cite{Nojiri:2015fra,Odintsov:2015zza,Oikonomou:2015qha,caruana2020cosmological}. This bouncing is generated by the scale factor 
\begin{eqnarray}
	a(t)=A_0e^{f_0\frac{(t-t_s)}{\alpha +1}^{\alpha +1}}\,,
	\label{eq:aIV}
\end{eqnarray}
where $A_0 > 0$ is a dimensionless constant that describes the scale factor at the bouncing time and $f_ 0>0$ is a constant with proper time dimension.
The parameter $\alpha$, if chosen equal to $1$ correspond to the symmetric bounce, that was discussed in Section \ref{sec:sym}. Moreover, the choice $\alpha = 0$ implies that the scale factor $a$ grows exponentially in time (de Sitter universe).

\begin{figure}[H]
	\centering 
	\begin{subfigure}[b]{0.45\textwidth}
		\centering
		\includegraphics[width=\textwidth,clip]{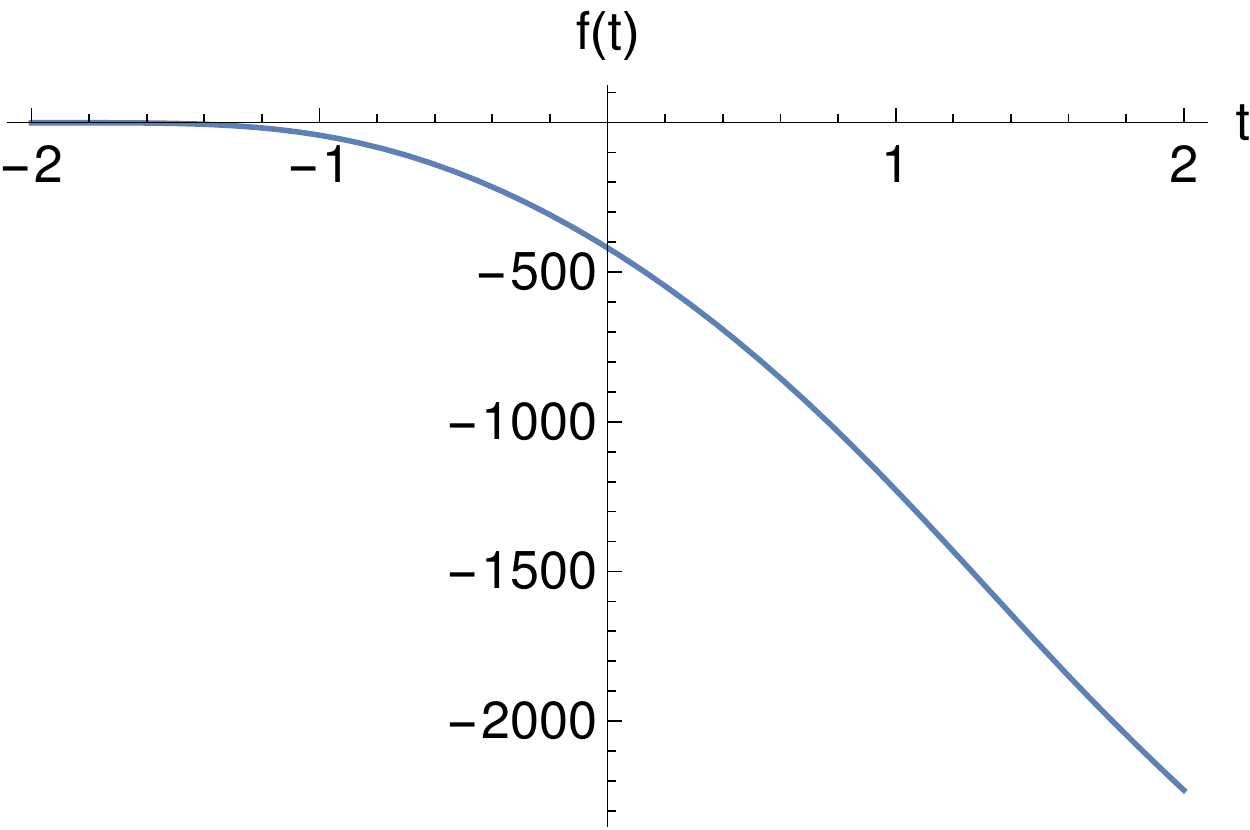}
		\caption{}
	\end{subfigure}
	\hfill
	\begin{subfigure}[b]{0.45\textwidth}
		\centering
		\includegraphics[width=\textwidth,clip]{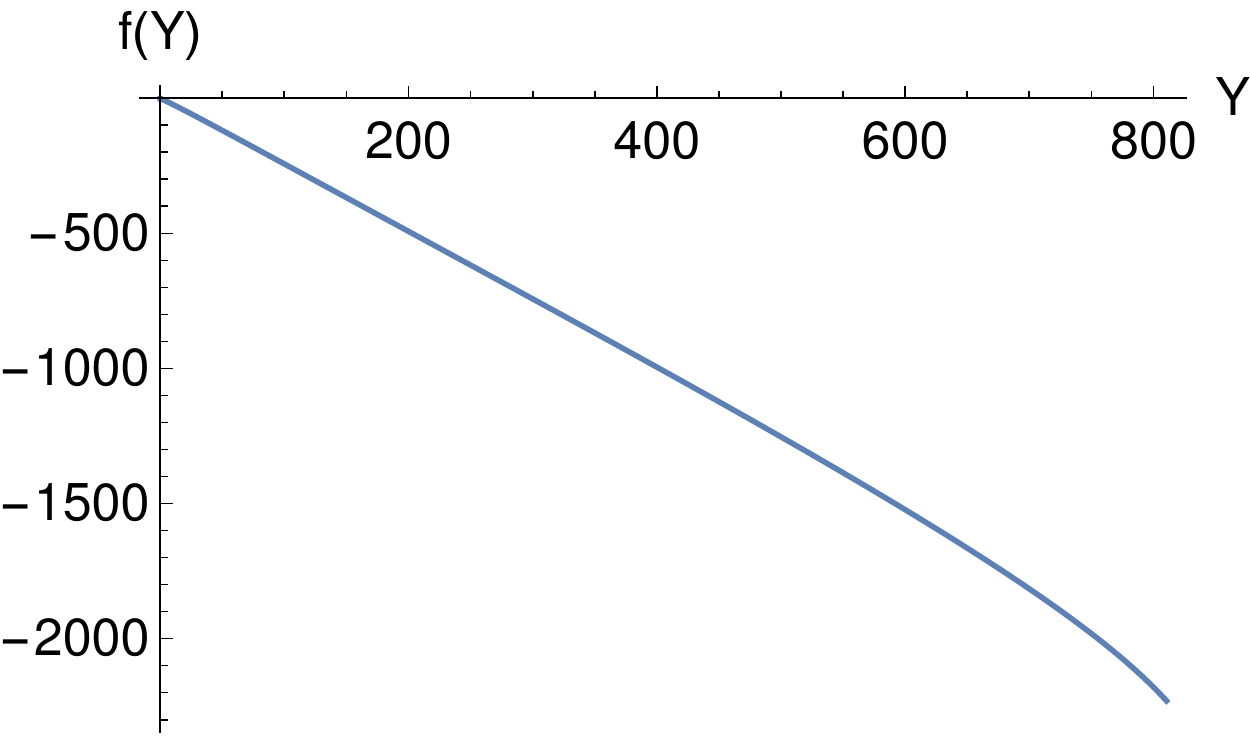}
		\caption{}
	\end{subfigure}
	\caption{\label{fig:ftIV} Numerical solution of the non-local distortion function in the finite time singularity model, panel (a) in terms of the cosmological time and panel (b) in terms of the field $Y$.}
\end{figure}

Now, we will consider the case $\alpha > 1$, such that, the scale factor and the effective energy density remains finite for every $t$. Panel (d) of the Figure \ref{fig:three graphs} shows the scale factor \eqref{eq:aIV} for the this model. Once again, the determination of $f(Y)$ follows the same procedure presented in the previous sections, although, only numerical solutions are available in this class of bounce. The resulting non-local distortion function for the $\alpha > 1$ case is presented in Figure \ref{fig:ftIV}, for the time dependence in panel (a) and for the non-local field $Y$ in panel (b). The function $f(t)$ is strictly negative, decay rapidly and goes to $-\infty$, when the present era is reached. The non-local distortion function is also strictly negative in this case and its behavior is nearly a straight line.

\subsubsection{Nonviable solutions}
\label{sec:non}

In addition, we have analyzed the solutions for the scale factor \eqref{eq:aIV} for $\alpha < -1$  which produces the Big Rip as a final fate of the universe, $0 < \alpha < 1$, which generates the sudden singularity model and for $-1 < \alpha < 0$, which results in the Big Freeze universe.
In the first scenario the scale factor diverges at the bouncing point, which causes divergences in the fields $X,Y,U$ and $V$ at this point and consequently the non-local modification in the DW II lagrangian density grows to infinity near the bouncing.
For the range $-1 < \alpha < 1$ is the first time derivative of the Hubble parameter that diverges at $t=0$, this causes a discontinuity in the non-local distortion function, presenting one solution for $t > 0$ and another for $t < 0$.
This discontinuity implies in an ambiguity in the DW II model, since the time do not enter in the action, therefore there is no way to chose which solution is the correct one.
Thus, we will regard these last cases as physically non-viable and infer that a well defined Hubble parameter and its derivative, as a necessary condition for the viability of the model.

\subsection{Pre-inflationary asymmetrical bounce}
\label{sec:pre}

Now we will study the non-local solutions in a pre-inflationary scenario for our Universe, recently proposed in the $f(R)$ modified gravity \cite{odintsov2022preinflationary}.
Specifically, in this scenario the Universe in the pre-inflationary era contracts until it reaches a minimum, and  expands slowly entering a quasi de Sitter inflationary era.
After that, the universe starts to contract again and the scale factor tends to zero.
The motivation to this suggestion is that it avoids the cosmic singularity and approximately satisfies the String Theory scale factor duality $a(t) = a^{-1}(-t)$.
The asymmetrical bouncing scale factor that describe this evolution is \cite{odintsov2022preinflationary}
\begin{eqnarray}
	a(t) = a_0e^{-H_b^3t^3-H_i^2t^2+H_0t}\,,
\end{eqnarray}
where $a_0$ is the minimum value of the scale factor, $H_b$, $H_i$ and $H_0$ are positive constants. 
We have presented this scale factor in panel (e) of Figure \ref{fig:three graphs}, note that we have chosen (by means of illustration) the bouncing point occurring at $t=-1$ and the quasi de Sitter expanding phase goes from $t=-0.5$ to $t=0.3$ for the numerical calculation purposes.

\begin{figure}[H]
	\centering 
	\begin{subfigure}[b]{0.45\textwidth}
		\centering
		\includegraphics[width=\textwidth,clip]{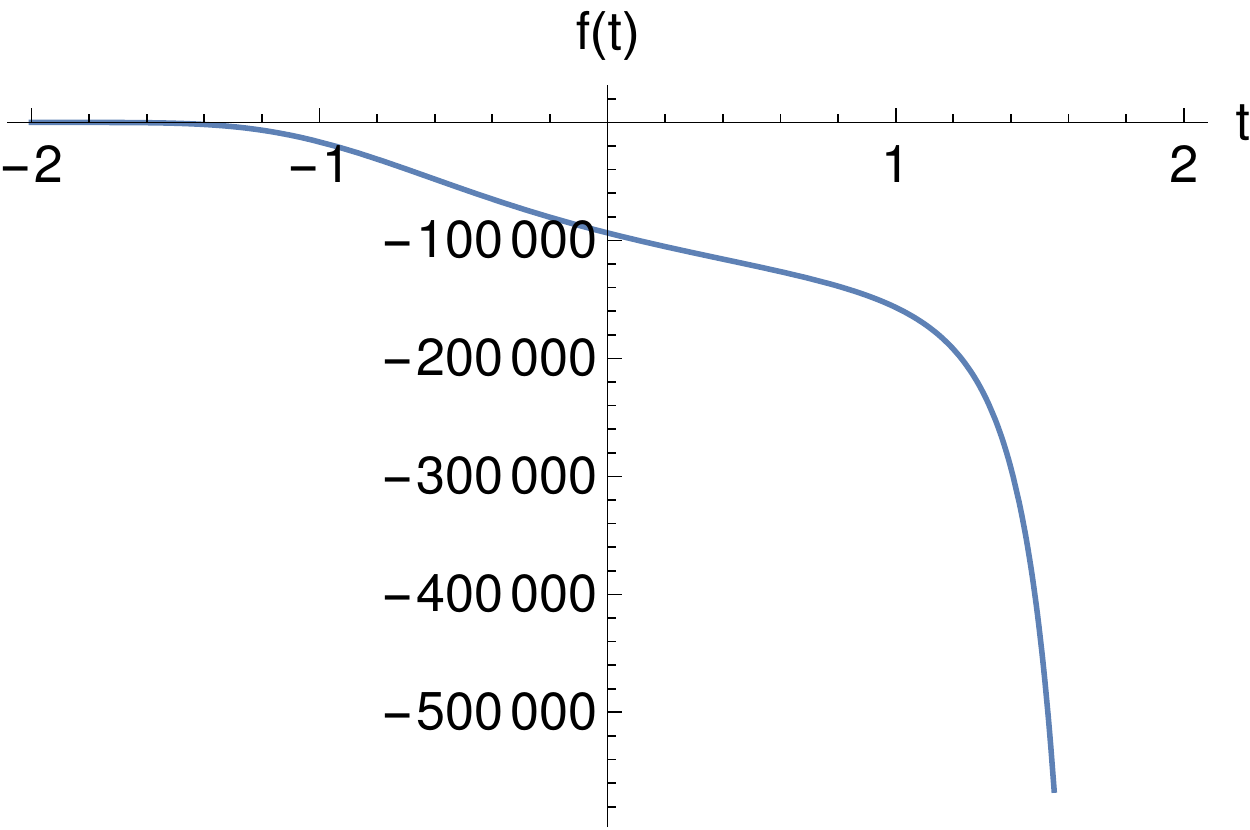}
		\caption{}
	\end{subfigure}
	\hfill
	\begin{subfigure}[b]{0.45\textwidth}
		\centering
		\includegraphics[width=\textwidth,clip]{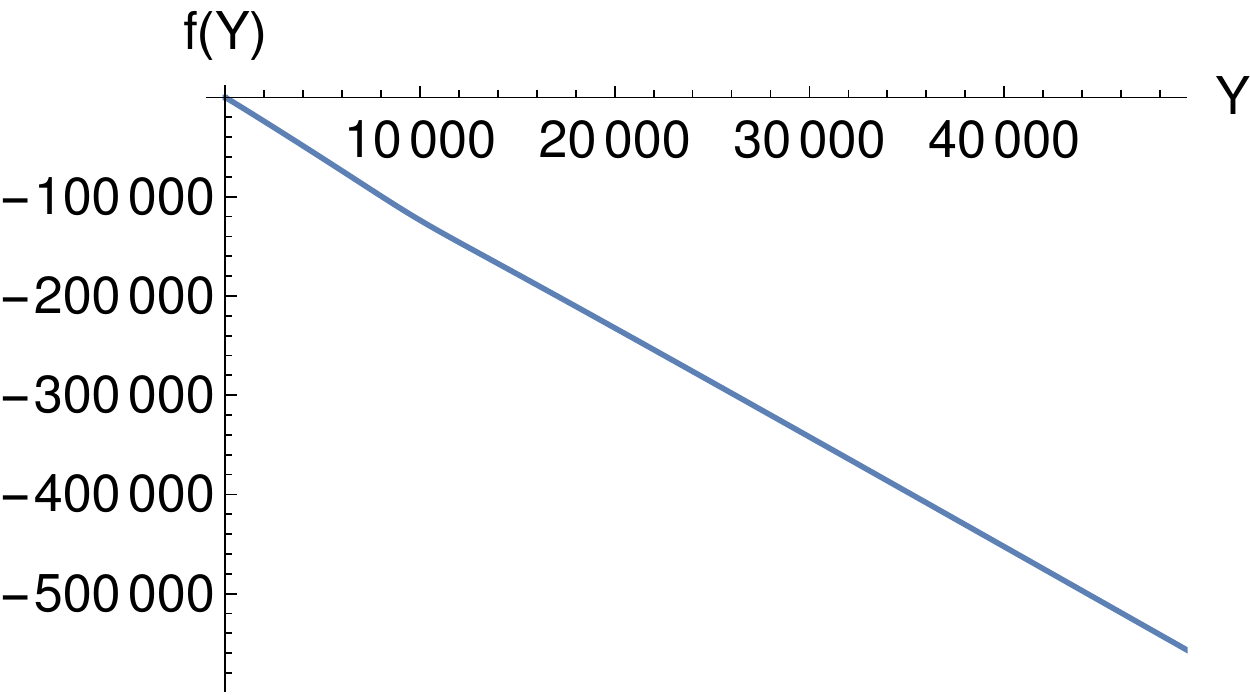}
		\caption{}
	\end{subfigure}
	\caption{\label{fig:ftV} Numerical solution of the non-local distortion function in the asymmetrical bounce, panel (a) in terms of the cosmological time and panel (b) in terms of the field $Y$.}
\end{figure}

The numerical solutions for the non-local distortion function $f(Y)$ and $f(t)$ are presented in Figure \ref{fig:ftV}.
In this case, the function $f(t)$ is also strictly negative and decreases almost linearly near the bouncing $t=-1$. Nonetheless, it is important to note that this behavior is kept during the quasi de Sitter expanding phase ($-0.5<t<0.3$).
When the contraction phase is reached again ($t>0.3$), the function starts to decay fast and goes to $-\infty$.
Furthermore, the non-local correction $f(Y)$ decreases linearly along all the evolution of the universe, without suffering any change over the contractions, expansion and re-contraction phases.

\section{Conclusions}
\label{sec:conc}

We have examined the viability of the improved Deser-Woodard non-local gravity model over the background of five different bouncing cosmologies and numerically reconstructed the non-local distortion function for each case.
The bouncing scenarios considered in our analysis have shown a physical and viable solution to the non-local distortion function $f(Y)$, namely: the symmetric bounce, the matter bounce, the finite time singularity model and the pre-inflationary asymmetrical bounce, with some exceptions in particular cases of finite time singularity. The distortion function $f(Y)$ encodes all the non-local information regarding the departure of the Einstein-Hilbert action and this kind of modified gravity theory is inspired by quantum phenomena.
Even though our reconstruction analysis is performed in a vacuum spacetime, we could expect that the inclusion of matter, for instance radiation, should not change our main results and conclusion (since the distortion function can also be obtained by solving the corresponding differential equations).

Although the improved Deser-Woodard model was initially proposed to reproduce the late time accelerated expansion of the universe (present in the $\Lambda$CDM model), we have found that this model can also describe the aforementioned bouncing scenarios, via the distortion function.
The solutions of $f(Y)$  that we have obtained by the reconstruction procedure, are presented in the Figures \ref{fig:ftI}, \ref{fig:ftII}, \ref{fig:ftIII}, \ref{fig:ftIV} and \ref{fig:ftV}.
The initial conditions used throughout the calculations are the retarded boundary conditions, which require that all the fields and its first time derivatives vanish in a initial value surface.

In regard to the physically admissible bouncing scenarios:

\begin{itemize}

\item In the analysis of the symmetric bounce scenario we have found that the non-local distortion function is an increasing function and passes throughout the bouncing point without changing its behavior.

\item The reconstructed distortion function for the oscillatory bounce is continuous, strictly positive and increases for every time throughout the evolution of the universe, but the growth speed slows down near the main bouncing.

\item Moreover, the matter bounce generates a distortion function that is strictly negative and grows linearly along the universe evolution.

\item The finite time singularity model, for $\alpha > 1$ have a $f(Y)$ solution that decays linearly for all $t$ in the solution interval considered.

\item Subsequently, the non-local correction generated in the pre-inflationary asymmetrical model also decreases linearly in the first contraction phase, the quasi de Sitter expansion and the second contraction phase, without suffering any change.

\end{itemize}

On the other hand, it is important to mention that the finite time singularity Big Rip model ($\alpha < -1$) presents divergences in the scale factor and in the Hubble parameter at the bouncing point $t=0$.
In the cases of the Big Freeze ($-1< \alpha < 0$) and sudden singularity model ($0 < \alpha < 1$), $a(t)$ and $H(t)$ are well defined, but the time derivative of the Hubble parameter $\dot{H}$ diverges at the origin.
These singularities cause the non-local fields $(X,Y,U,V)$ to diverge and also the distortion function to grows to $\pm \infty$ near the bouncing.
In a closer look, all these scenarios present a Hubble parameter or its time derivative not well defined at $t=0$, this causes a discontinuity in the non-local distortion function and two solutions (one for $t>0$ and another for $t<0$) must be considered. Since the time do not enter in the lagrangian, there is no way to choose what is the right solution for $f(Y)$, hence, this causes an ambiguity in the model. Hence, based in these remarks, we can ascribed the physical non-viability of these bouncing models, in the context of the Deser-Woodard cosmology, to the singular nature of the Hubble parameter. A well defined Hubble parameter (and its derivative) over all the interval of solution can be seen as a strong constraint as a first criteria in the search for viable bouncing scenarios. But this statement requires further analysis to examine its generality.

We believe that the present analysis and physical results of bouncing solutions (at the level of background cosmology in the DW II gravity)  motivate a deeper look at the early Universe cosmology.
To further restrict physically relevant models, we need to study the early time perturbations of these bouncing models and examine its implications, for instance, on the cosmic microwave background.
This and other topics are now under development and will be reported elsewhere.
 

\acknowledgments

The authors would like to thank the anonymous referee for his/her comments and suggestions to improve this paper.
This study was financed in part by the Coordenação de Aperfeiçoamento de Pessoal de Nível Superior - Brasil (CAPES) - Finance Code 001.
R.B. acknowledges partial support from Conselho
Nacional de Desenvolvimento Cient\'ifico e Tecnol\'ogico (CNPq Projects No. 305427/2019-9 and No. 421886/2018-8) and Funda\c{c}\~ao de Amparo \`a Pesquisa do Estado de Minas Gerais (FAPEMIG Project No. APQ-01142-17).



\begin{thebibliography}{99}

\bibitem{will2014confrontation}
C. M. Will, \emph{The confrontation between general relativity and experiment}, \emph{Living Rev. Relativ.} {\bf vol 17} (2014).

\bibitem{hawking1973large}
S. W. Hawking and G. F. R. Ellis, 
\emph{The large scale structure of space-time},
{\bf vol 1},
 Cambridge University Press (1973).

\bibitem{misner1973gravitation}
C. W. Misner, K. S. Thorne, and J. A. Wheeler, \emph{Gravitation},
Macmillan (1973).

\bibitem{geroch1968singularity}
R. Geroch, \emph{What is a singularity in general relativity?} \emph{Ann. of Phys.}, {\bf vol 17}, 3, Elsevier, (1968), pp. 526--540.


\bibitem{Bojowald:2001xe}
M. Bojowald,
\emph{Absence of singularity in loop quantum cosmology},
	\emph{Phys. Rev. Lett.},
	{\bf vol 86},
	pg. 5227-5230,
	APS, (2001).
			
\bibitem{novello2008bouncing}
M. Novello  and S. E. Perez Bergliaffa,
	\emph{Bouncing Cosmologies},
	\emph{Phys. Reports},
	{\bf vol 463},
	pg. 127--213,
	Elsevier, (2008).

\bibitem{sachs1996thequasi}
R. Sachs, J. V. Narlikar and F. Hoyle.,
	\emph{The quasi-steady state cosmology: analytical solutions of field equations and their relationship to observations.},
	\emph{Astr. and Astrophys.},
	{\bf vol 313},
	pg 703-712,
	SAO/NASA, (1996).

\bibitem{cai2012bounce}
Y. Cai, C. Gao, and E. N. Saridakis,
	\emph{bounce and cyclic cosmology in extended nonlinear massive gravity},
	\emph{Jour.of Cosmo. and Astr. Phys.},
	IOP Publishing,
	(2012).

\bibitem{mukherjo2002bouncing}
S. Mukherji, and M. Peloso,
	\emph{Bouncing and cyclic universes from brane models},
    \emph{Phys. Lett. B},
	Elsevier,
	(2002).


\bibitem{Ashtekar:2011ni}
A.~Ashtekar and P.~Singh,
\emph{Loop Quantum Cosmology: A Status Report},
	\emph{Class. Quant. Grav.},
	{\bf vol 28},
	pg. 213001,
	IOP, (2011).
	
	

	\bibitem{battefeld2015critical}
D. Battefeld, and P. Peter,
	\emph{A critical review of classical bouncing cosmologies},
	\emph{Phys. Reports},
	{\bf vol 571}
	pg. 1--66,
	Elsevier,
	(2015).

\bibitem{Brandenberger:2017ni}
R. Brandenberger and P. Peter,
\emph{Bouncing Cosmologies: Progress and Problems},
	\emph{Found. Phys.},
	{\bf vol 47},
	pg. 797-850,
	Springer, (2017).

\bibitem{Nojiri:2017ni}
S. Nojiri, S. D. Odintsov, and V. K. Oikonomou,
\emph{Modified Gravity Theories on a Nutshell: Inflation, bounce
and Late-time Evolution},
	\emph{Phys. Rept.},
	{\bf vol 692},
	pg. 1-104,
	Elsevier, (2017).
	
	
\bibitem{mukhanov:2005rb}
V. Mukhanov,
\emph{Physical Foundations of Cosmology},
Cambridge University Press, Oxford, 2005.



\bibitem{odintsov:2015ni}
K. Bamba and S. D. Odintsov,
\emph{Inflationary cosmology in modified gravity theories},
	\emph{Symmetry},
	{\bf vol 7},
	pg. 220-240,
	MDPI, (2015).
	
	
\bibitem{ishak2019testing}
M. Ishak, \emph{Testing general relativity in cosmology}, \emph{Living Rev. Relativ.},
	{\bf vol 22},
	1,
	Springer,
	(2019).
	
\bibitem{belgacem2018nonlocal}
	E. Belgacem, Y. Dirian, S. Foffa, and M. Maggiore,
	\emph{Nonlocal gravity. Conceptual aspects and cosmological predictions},
	\emph{J. of Cosmo. and Astr. Phys.},
	{\bf vol 03},
	IOP Publishing,	(2018).
	
\bibitem{Capozziello:2022lic}
S.~Capozziello and F.~Bajardi,
\emph{Non-Local Gravity Cosmology: an Overview},
arXiv:2201.04512 [gr-qc].
	
\bibitem{dalvit1994running}
	D. A. R. Dalvit and F. D. Mazzitelli,
	\emph{Running coupling constants, Newtonian potential, and non-localities in the effective action},
	\emph{Phys. Rev. D},
	{\bf vol 50},
	2,
	pg. 1001, APS,
	(1994).

\bibitem{wetterich1998effective}
	C. Wetterich,
	\emph{Effective non-local Euclidean gravity},
	\emph{Gen. Relativ. and Gravit.},
	{\bf vol 30},
	1,
	pg. 159--172,
	Springer,
	(1998).
	
\bibitem{Barvinsky:1985rb}	
A. Barvinsky and G. Vilkovisky,
\emph{The Generalized Schwinger-DeWitt Technique in Gauge
Theories and Quantum Gravity,}
\emph{Phys. Rept.},
	{\bf vol 119},
	pg. 1-74,
	Elsevier, (1985).
	
\bibitem{Buchbinder:1992rb}
 I.~L. Buchbinder, S. D. Odintsov, and I. L. Shapiro,
\emph{Effective action in quantum gravity},
Bristol, UK: IOP (1992).


\bibitem{Mukhanov:2007rb}
V. Mukhanov and S. Winitzki, 
\emph{Introduction to quantum effects in gravity},
University Press, Cambridge, (2007).
 
\bibitem{Shapiro:2008sf}
I.~L.~Shapiro,
{\em Effective Action of Vacuum: Semiclassical Approach},
 \emph{Class. Quant. Grav.},
	{\bf vol 25},
	pg. 103001,
	IOP, (2008).


\bibitem{maggiore2014nonlocal}
	M. Maggiore and M. Mancarella,
	\emph{Nonlocal gravity and dark energy},
	\emph{Phys. Rev. D},
	{\bf vol 90}, 2, APS,
		(2014).
\bibitem{barvinsky2012serendipitous}
A. Barvinsky,
	\emph{Serendipitous discoveries in nonlocal gravity theory},
	\emph{Phys. Rev. D},
	{\bf vol 85},
	10,
	APS,
	(2012).
	
\bibitem{deser2007nonlocal}
	S. Deser and R. P. Woodard,
	\emph{Nonlocal cosmology},
	\emph{Phys. Rev. Lett.},
	{\bf vol 99},
	11,
	APS, (2007).
	

\bibitem{deser2019nonlocal}
	S. Deser and R. P. Woodard,
	\emph{Nonlocal cosmology {II}. Cosmic acceleration without fine tuning or dark energy},
	\emph{J. of Cosmo. and Astr. Phys.},
	{\bf vol 06},
	pg. 34,
	IOP Publishing, (2019).
	
	

\bibitem{caruana2020cosmological}
M. Caruana,  G. Farrugia, and J. L. Said. 
	\emph{Cosmological bouncing solutions in f (T, B) gravity} 
	\emph{Eur. Phys. J. C} 80.7
	pg. 1-20 
	Springer, (2020).
	
\bibitem{chen2019primordial}
C.~Y.~Chen, P.~Chen and S.~Park,
	\emph{Primordial bouncing cosmology in the {D}eser-{W}oodard nonlocal gravity},
	\emph{Phys. Lett. B},
	{\bf vol 796},
	pg. 112--116,
	Elsevier, (2019).

	  
\bibitem{Cai:2012va}
Y.~F.~Cai, D.~A.~Easson and R.~Brandenberger,
\emph{Towards a Nonsingular Bouncing Cosmology},
JCAP \textbf{08} (2012), 020

\bibitem{Cai:2013vm}
Y.~F.~Cai, R.~Brandenberger and P.~Peter,
\emph{Anisotropy in a Nonsingular bounce},
Class. Quant. Grav. \textbf{30} (2013), 075019

	\bibitem{Bamba:2014mya}
K.~Bamba, A.~N.~Makarenko, A.~N.~Myagky and S.~D.~Odintsov,
\emph{Bouncing cosmology in modified Gauss-Bonnet gravity,}
Phys. Lett. B \textbf{732} (2014), 349-355



\bibitem{Steinhardt:2001st}
P.~J.~Steinhardt and N.~Turok,
\emph{Cosmic evolution in a cyclic universe},
Phys. Rev. D \textbf{65} (2002), 126003

\bibitem{Singh:2006im}
P.~Singh, K.~Vandersloot and G.~V.~Vereshchagin,
\emph{Non-singular bouncing universes in loop quantum cosmology},
Phys. Rev. D \textbf{74} (2006), 043510

\bibitem{Wilson-Ewing:2012lmx}
E.~Wilson-Ewing,
\emph{The matter bounce Scenario in Loop Quantum Cosmology},
JCAP \textbf{03} (2013), 026

\bibitem{Barrow:2015ora}
J.~D.~Barrow and A.~A.~H.~Graham,
\emph{Singular Inflation},
Phys. Rev. D \textbf{91} (2015) no.8, 083513

\bibitem{Nojiri:2015fra}
S.~Nojiri, S.~D.~Odintsov and V.~K.~Oikonomou,
\emph{Quantitative analysis of singular inflation with scalar-tensor and modified gravity},
Phys. Rev. D \textbf{91} (2015) no.8, 084059

\bibitem{Odintsov:2015zza}
S.~D.~Odintsov and V.~K.~Oikonomou,
\emph{Bouncing cosmology with future singularity from modified gravity},
Phys. Rev. D \textbf{92} (2015) no.2, 024016

 \bibitem{Oikonomou:2015qha}
 V.~K.~Oikonomou,
\emph{Singular Bouncing Cosmology from Gauss-Bonnet Modified Gravity},
Phys. Rev. D \textbf{92} (2015) no.12, 124027

\bibitem{odintsov2022preinflationary}
S. D. Odintsov and V. K. Oikonomou,
\emph{Pre-inflationary bounce effects on primordial gravitational waves of $f(R)$ gravity},
\emph{Phys. Lett. B},
{\bf vol 824}, 136817 (2022).

\bibitem{ding2019structure}
J.~C.~Ding and J.~B.~Deng,
	\emph{Structure formation in the new {D}eser-{W}oodard nonlocal gravity model},
	\emph{J. of Cosmo. and Astr. Phys.},
	{\bf vol 12},
	pg. 054,
	{IOP} Publishing,
	(2019).
	


\end{thebibliography}
\end{document}